\documentclass[12pt,letterpaper]{article}
\usepackage{amsmath,amssymb,calc}
\usepackage{graphicx}
\usepackage{color}

\newcommand\be{\begin{equation}}
\newcommand\ee{\end{equation}}
\newcommand{\bea}{\begin{eqnarray}}
\newcommand{\eea}{\end{eqnarray}}

\newcommand{\nn}{\nonumber}
\newcommand{\pd}{\partial}

\def\id{\protect{{1 \kern-.28em {\rm l}}}}

\def\1{^{(1)}}
\def\0{^{(0)}}
\def\2{^{(2)}}

\def\id{\protect{{1 \kern-.28em {\rm l}}}}
\let\non\nonumber

\begin{document}
\begin{titlepage}

\begin{center}
\hfill \\
%\hfill {\tt ...}\\
%\vskip 15mm

%\vskip 2cm
\vskip 1cm 
%{\Large \bf Holographic Walking Technicolor\\ \vspace{0.4cm}\\from D-branes}
{\Large \bf Holographic Walking Technicolor \vspace{0.4cm}\\from D-branes}

\vskip 1.5 cm 
{\bf  Lilia Anguelova\footnote{anguella@ucmail.uc.edu}, Peter Suranyi\footnote{peter.suranyi@gmail.com} and L.C.R. Wijewardhana\footnote{rohana.wijewardhana@gmail.com}}\non\\
{\vskip 0.5cm  {\it Dept. of Physics, University of Cincinnati,
Cincinnati, OH 45221, USA}\non\\}

\end{center}
\vskip 2 cm

\begin{abstract}
\baselineskip=18pt

We investigate a model of dynamical electroweak symmetry breaking via a dual gravitational description. The gravity dual is obtained by embedding a D7-$\overline{{\rm D}7}$ pair of branes into a type IIB background that is dual to a walking gauge theory. We develop further a previous study of this model. In particular, we show that there is a nontrivial relation that needs to be satisfied in order for axial-vector modes to exist. Furthermore, we compute explicitly the electroweak $S$ parameter. The result is positive-definite and, as was to be expected, much smaller than in earlier QCD-like D-brane constructions. We also find the masses and decay constants of the vector and axial-vector mesons in this model. This allows us to obtain another estimate for $S$ by summing the contributions of the discrete states. It is noteworthy that, in contrast to previous holographic studies, the sum of the first several lowest-lying states {\it does} give a very good approximation to the full answer.

\end{abstract}

\end{titlepage}

\tableofcontents

\section{Introduction}

The spontaneous breaking of electroweak symmetry and the ensuing generation of gauge boson masses are key ingredients of modern particle physics. Uncovering the origin of this electroweak symmetry breaking (EWSB) is one of the most important goals of future experiments. In the Standard Model the mass generation is due to the presence of an $SU(2)$ doublet Higgs field, comprised of four real components. Spontaneous EWSB occurs when this field acquires a non-vanishing vacuum expectation value. Three of its components then give rise to the would-be Nambu-Goldstone bosons of the spontaneously broken symmetry, which become the longitudinal third polarizations of the massive  $W^{\pm}$ and $Z$ bosons. The remaining scalar degree of freedom is the Higgs boson. 

Notably, however, the Higgs boson has not yet been discovered experimentally. Furthermore, as it is a fundamental scalar, its mass diverges due to quantum corrections. This problem can be significantly alleviated in supersymmetric extensions of the Standard Model. Namely, supersymmetry (SUSY) tames the quantum corrections to the Higgs mass by providing automatic cancellations between bosonic and fermionic superpartner contributions. Despite that however, an initial fine-tuning is still necessary. This lack of understanding of the hierarchy between the electroweak (Higgs mass) scale and the cutoff (Planck) scale is known as unnaturalness problem. In addition, the question of SUSY breaking remains to be settled. 

All those reasons have motivated the search for a more conceptually appealing alternative. A candidate for such an alternative, that leads to EWSB without any fine-tuning, is offered by technicolor (TC) theories \cite{Tech}. In the latter, instead of the Higgs mechanism, one uses dynamical chiral symmetry breaking in a new strongly-coupled gauge sector. Initially, TC models were thought to have dynamics similar to QCD, only at a higher energy scale. However, such TC theories face important problems when compared to phenomenology. Namely, they predict too large a value for the Peskin-Takeuchi S-parameter \cite{PT}. Furthermore, the introduction of extended technicolor (ETC) interactions \cite{ETC}, in order to generate fermion masses, leads to unacceptably large flavor-changing neutral current amplitudes.

A phenomenologically promising modification of the above ideas is provided by walking TC (WTC) theories \cite{WalkTech}. A characteristic feature of the latter is that the gauge coupling runs much slower than that in QCD between the TC and ETC scales. This is due to a different field content compared to that in QCD-like models. The presence of such a walking (almost conformal) intermediate region is expected to help solve the problem with the magnitude of the flavor-changing neutral currents. Starting from \cite{IRfpb}, vector gauge theories, that exhibit strong IR fixed point behavior, have been used to build WTC models.\footnote{In \cite{ST} the techniquarks were placed in higher representations, in order to achieve walking with a smaller techniflavor symmetry group and thus to reduce the value of $S$.} It has also been conjectured that WTC theories should predict smaller S-parameter than QCD-like TC ones, due to the markedly different bound state spectra \cite{KL,AS}. Recall that, in QCD-like technicolor the S-parameter was obtained by scaling-up to the TC scale the measured spectral functions of QCD. In WTC, however, such an estimate is clearly incorrect. Instead, one has to address the computation anew, while working at strong coupling. This is, obviously, beyond the realm of perturbative QFT methods and thus is the main issue that has long hindered the investigation of walking technicolor. For reviews of WTC, see \cite{MP}.\footnote{We should note that in recent years nonperturbative methods, like lattice simulations \cite{LSD} and solving bound state equations \cite{KSY}, have been utilized to estimate the spectra of non-QCD-like theories. These investigations show reduction in $S$ compared to QCD-like models.}

Modern developments in gauge/gravity duality provide exactly the tools to address the above problem, namely to study gauge theories at strong coupling. What we mean by this is the following. Gauge theories, that can be realized as the worldvolume degrees of freedom of D-brane configurations in string theory, have dual descriptions in terms of the gravitational backgrounds created by those branes. Most importantly, the strongly-coupled regime of the field theory corresponds to the weakly-coupled regime of the gravity dual. Hence, one can use classical supergravity to obtain answers to questions about strongly-coupled gauge theory. This remarkable duality has been utilized to investigate a wide-ranging set of problems, from hydrodynamics \cite{hydro} to superconductivity \cite{scond}.\footnote{It should be noted though that there is a large variety of AdS/CFT-inspired phenomenological models, whose string theory embedding is an open problem.} Most relevant for us, \cite{SS,SS2} built a gravitational dual of a gauge theory that exhibits many of the characteristic features of QCD. In this holographic QCD model, chiral symmetry is spontaneously broken due to a particular U-shaped embedding of flavour D8-$\overline{{\rm D}8}$ pairs into a background sourced by color D4-branes.

The above Sakai-Sugimoto model can easily be adopted to give a gravitational dual of QCD-like technicolor. The latter, as well as other similar models obtained from various brane configurations in type IIA/B string theory, has been studied in \cite{HolTech}.\footnote{Note that there is a huge, (loosely) inspired by AdS/CFT, literature on holographic technicolor models \cite{PhenoHolTech}, which do not have a consistent string theory embedding.} With this method one can compute explicitly the Peskin-Takeuchi $S$, $T$ and $U$ parameters \cite{PT}, that are important electroweak observables on a par with the fine structure constant and the mass of the Z-boson. As is to be expected though, the QCD-like technicolor models of \cite{HolTech} are incompatible with present observational bounds. In \cite{LA}, a gravitational dual of a walking technicolor model was considered. A key ingredient was the type IIB background of \cite{NPP}, which is dual to a walking gauge theory. One can show that there is a U-shaped embedding, a la Sakai-Sugimoto, of a probe D7-$\overline{{\rm D}7}$ pair in that background \cite{LA}. Thus, one can realize geometrically  chiral symmetry breaking, which then induces electroweak symmetry breaking.

Here we will study further this gravity dual of walking techinicolor. First, we will show that there is a non-trivial constraint between the three parameters of the model, which has to be satisfied in order for regular axial-vector modes to exist. Hence, the number of free parameters is with one less than initially expected. Then we will compute analytically the $S$ parameter via the exact expression in terms of non-normalizable modes derived in \cite{HolTech, LA}. Unlike \cite{LA} however, our present method for finding the non-normalizable solutions will not leave any undetermined constants. The explicit answer we obtain for $S$ is positive-definite and compatible with the current experimental bounds, although it does exhibit an unexpected functional dependence on the length of the walking region. We also compute the masses and decay constants for the spectrum of vector and axial-vector mesons in this model. This allows us to obtain yet another estimate for the value of the $S$ parameter, by summing the contributions of the discrete states. The result agrees completely with the answer from the non-normalizable modes, in contrast to the QCD-like technicolor models of \cite{HolTech}. In fact, the sum over discrete states converges rather rapidly. Hence, summing several lowest lying states does give a very good approximation to the full answer. Finally, it is worth noting that we can relate directly to each other, i.e. without any of the model parameters present, the physical observables $S$, $F_{\pi}$ and $m_n^V$, where $F_{\pi}$ is the technipion decay constant and $m_n^V$ is the mass of the $n^{\rm th}$ vector state. This relation implies, in particular, that for a single techniflavour pair the lightest vector meson mass $m_1^V \ge 5$ TeV, in order for the experimental bound on the $S$ parameter to be satisfied.

\section{Gravitational dual}
\setcounter{equation}{0}

One can build gravitational duals of technicolor models in the following manner. Let us consider a background sourced by a stack of $N_{TC}$ technicolor branes in type IIA/B string theory. To add techniflavour degrees of freedom, we need to include an additional set of $N_{TF}$ probe branes and $N_{TF}$ probe antibranes. If there is a U-shape embedding a la Sakai-Sugimoto \cite{SS} of the techniflavour probes into the technicolor backrgound, such that the probe branes and antibranes merge at some finite radial distance, then we have chiral symmetry breaking. This is because the techniflavour group $U_L (N_{TF}) \times U_R (N_{TF})$, arising from the separate branes and antibranes, gets broken to the diagonal subgroup $U (N_{TF})$ due to the merging. This chiral symmetry breaking then induces electroweak symmetry breaking via an appropriate choice of embedding of the electroweak $SU(2)\times U(1)$ group into the techniflavour $U_L (N_{TF}) \times U_R (N_{TF})$. The above general set up has been utilized to study a variety of QCD-like technicolor models \cite{HolTech}. In \cite{LA} it was applied to build a model of walking technicolor. An essential ingredient in that regard is the recently discovered gravitational background dual to walking behaviour \cite{NPP}.\footnote{We should note that this is within the class of the solutions studied in \cite{CNP}.} In this section we review briefly the background of \cite{NPP} and then discuss in more detail the U-shaped techniflavour probe embedding studied in \cite{LA}. Finally, we show that there is a nontrivial condition, that has to be satisfied in order for regular axial-vector modes to exist in this model.

\subsection{Walking technicolor background} 

The background of \cite{NPP} is an ${\cal N} = 1$ solution of the type IIB equations of motion with nontrivial RR 3-form flux and constant dilaton. It is a deformation of the Maldacena-Nunez background \cite{MN}, that still arises from D5-branes wrapping an $S^2$. The ten-dimensional metric is:
\bea \label{BM}
ds^2 &=& A \left[ dx_{1,3}^2 + \frac{cP_1' (\rho)}{8} \left( 4 d\rho^2 + (\omega_3 + \tilde{\omega}_3)^2 \right) \right. \nn \\
&+& \left. \frac{c\,P_1 (\rho) \coth (2 \rho)}{4} \left( d\Omega_2^2 + d\tilde{\Omega}_2^2 + \frac{2}{\cosh ( 2\rho)} (\omega_1 \tilde{\omega}_1 - \omega_2 \tilde{\omega}_2) \right) \right] \, ,
\eea
where\footnote{In the expression for $A$ we have corrected a typo in \cite{NPP}.}
\be \label{AP1}
A=\left( \frac{3}{c^3 \sin^3 \alpha} \right)^{1/4} , \quad  P_1' (\rho) =\frac{\pd P_1 (\rho)}{\pd \rho} \,\, , \quad P_1 (\rho) = \left( \cos^3 \alpha + \sin^3 \alpha \left( \sinh (4 \rho) - 4 \rho \right) \right)^{1/3} \, ,
%P_1' (\rho)& = &\frac{8}{3}\,\beta \,\sinh^2(2\,\rho)\,[P_1(\rho)]^{-2}.
\ee
\bea
\tilde{\omega}_1 &=& \cos \psi d\tilde{\theta} + \sin \psi \sin \tilde{\theta} d \tilde{\varphi} \,\, , \hspace*{2cm} \omega_1 = d \theta \,\, , \nn \\
\tilde{\omega}_2 &=& - \sin \psi d\tilde{\theta} + \cos \psi \sin \tilde{\theta} d \tilde{\varphi} \,\, , \hspace*{1.6cm} \omega_2 = \sin \theta  d \varphi \,\, , \nn \\
\tilde{\omega}_3 &=& d \psi + \cos \tilde{\theta} d \tilde{\varphi} \,\, , \hspace*{3.7cm} \omega_3 = \cos \theta d \varphi
\eea
and
\be
d\tilde{\Omega}_2^2 = \tilde{\omega}_1^2 + \tilde{\omega}_2^2 \,\, , \hspace*{2cm} d\Omega_2^2 = \omega_1^2 + \omega_2^2 = d \theta^2 + \sin^2 \theta d \varphi^2 \,\, .
\ee
Here the constants $c$ and $\alpha$ are parameters characterising the solution and, as in \cite{NPP}, we have set $\alpha' = 1 = g_s$. Also, we will not write down the explicit form of the RR 3-form flux, since it will not be needed for our purposes.

The walking region corresponds to the interval $\rho \in (1, \rho_*)$, where the upper end is approximately given by\footnote{We will comment more on the value of $\rho_*$ in Section \ref{Sexact}.} 
\be \label{cutoffNPP}
\rho_* \approx \frac{1}{4} \left( \log 2 + 3 \log \cot \alpha \right) \, .
\ee
This can be seen by investigating the dependence of the technicolor gauge coupling $g^2$ on the radial variable $\rho$, which can be read off from the DBI action of a supersymmetric D5-brane probe in the above background \cite{NPP}:
\be \label{g2TC}
\frac{g^2}{8 \pi^2} = \frac{e^{{\rm arcsinh} \left(\frac{1}{\sinh (2 \rho)} \right)}}{cP_1} = \frac{e^{{\rm arcsinh} \left(\frac{1}{\sinh (2 \rho)} \right)}}{c \left( \cos^3 \alpha + \sin^3 \alpha \left( \sinh (4 \rho) - 4 \rho \right) \right)^{1/3} } \,\,\, .
\ee
Let us also note that in the walking region the metric (\ref{BM}) simplifies to:
\be \label{simpM}
ds^2 \approx A \left[ dx^2_{1,3} + \frac{c \cos \alpha}{4} \left( \frac{\tan^3 \alpha \,e^{4 \rho}}{3} \left( 4 d \rho^2 + (\omega_3 + \tilde{\omega}_3)^2 \right) + d\Omega^2_2 + d\tilde{\Omega}^2_2 \right) \right].
\ee
Finally, the validity of the approximation, in which the above solution has been found, requires that the parameter $c>\!\!>1$, whereas for the existence of a walking region one needs that $\beta \equiv \sin^3 \alpha <\!\!<1$.

\subsection{Techniflavour probe branes} 

To add techniflavour degrees of freedom, we will consider probe D7 branes in the background (\ref{BM}). In \cite{LA} it was shown that one can find a probe embedding of the necessary U-shape type for the following ansatz: The D7 worldvolume directions are taken to be the four space-time dimensions $x^{\mu}$, the radial variable $\rho$ and the 3-sphere parameterized by $\psi$, $\tilde{\theta}$ and $\tilde{\varphi}$. Furthermore, the transverse coordinates $\theta$ and $\varphi$ are functions of $\rho$ only, i.e. $\theta = \theta (\rho)$ and $\varphi = \varphi (\rho)$.\footnote{This is very similar to the embedding of D7 probes in the conifold, considered in \cite{KS}.}

The embedding functions $\theta (\rho)$ and $\varphi (\rho)$ are determined by the field equations that follow from the DBI action for the probe branes or, in other words, from the Lagrangian
\be
{\cal L} = const \sqrt{- \det (g_8)} \,\, , 
\ee 
where $g_8$ is the metric induced on the D7-brane world-volume, i.e. the metric (\ref{BM}) with $\theta = \theta (\rho)$ and $\varphi = \varphi (\rho)$ substituted. More explicitly, we have
\be \label{Lagr}
\sqrt{- \det (g_8)} = \left[ f(\rho) + g(\rho) \left( \theta_{\rho}^2 + \sin^2 \theta \varphi_{\rho}^2 \right) \right]^{1/2} \,\, ,
\ee 
where $\theta_{\rho} = \pd \theta / \pd \rho$ , \,$\varphi_{\rho} = \pd \varphi / \pd \rho$ \,and
\bea \label{fg}
f(\rho) &=& \frac{1}{256} \times A^8 c^4 P_1^2 (\rho) P_1'^2 (\rho) \coth^2 (2 \rho) \nn \\
g(\rho) &=& \frac{1}{256} \times \frac{A^8 c^4}{2} P_1^3 (\rho) P_1' (\rho)\coth (2 \rho) \,\, .
\eea
It is easy to see that the equation of motion for $\theta (\rho)$ is solved by $\theta = \pi / 2$ for every $\rho$, just like in the simplified walking metric (\ref{simpM}) that was studied in \cite{LA}. Then the Euler-Lagrange equation for $\varphi (\rho)$, that follows from the Lagrangian (\ref{Lagr}), is:
\be \label{EL}
\frac{\pd }{\pd \rho} \,\frac{d {\cal L}}{ d \varphi_{\rho}} = 0 \, ,
\ee
which implies that:
\be \label{CQ}
\frac{d {\cal L}}{ d \varphi_{\rho}} = C 
\ee
with $C$ being a constant. More explicitly, (\ref{CQ}) has the form 
$\frac{g (\rho) \varphi_{\rho}}{\sqrt{f (\rho) + g (\rho) \varphi^2_{\rho}}} = C$,
which is just an algebraic equation for $\varphi_{\rho}$ that can easily be solved:
\be \label{Solrho0}
\varphi_{\rho}^2 = \frac{C^2 f (\rho)}{g (\rho) \left[ g(\rho) - C^2 \right]} \, .
\ee

To fix the constant $C$, let us look at how things simplify in the walking region. Due to the simplified metric (\ref{simpM}), the Lagrangian acquires the form:
\be
{\cal L} = const \times e^{2 \rho} \!\left( 4 B e^{4 \rho} + \varphi_{\rho}^2 \right)^{1/2} \, ,
\ee
where $B = \frac{\tan^3 \alpha}{3}$. Hence, the general equation (\ref{CQ}) reduces to
$\frac{e^{2 \rho} \varphi_{\rho}}{\left( 4B e^{4 \rho} + \varphi_{\rho}^2 \right)^{1/2}} = C$,
which implies that
\be
\varphi_{\rho} = \pm 2 \sqrt{B} C \frac{1}{\left( 1- \frac{C^2}{e^{4 \rho}} \right)^{1/2}} \,\, .
\ee
Comparing this with the walking region embedding equation (4.19) of \cite{LA}, we conclude that:
\be \label{Crho0}
C = e^{2 \rho_0} \, ,
\ee
where $\rho_0$ is the value of $\rho$ at which the D7 and anti-D7 branes merge, i.e. the tip of the U-shape embedding. To recapitulate, the embedding function $\varphi (\rho)$ is determined by
\be \label{VPrho0}
\varphi_{\rho}^2 = \frac{e^{4 \rho_0} f (\rho)}{g (\rho) \left[ g(\rho) - e^{4 \rho_0} \right]} \, .
\ee

So far, we have only addressed the embedding of the techniflavour branes into the background (\ref{BM}). In order to study vector and axial-vector mesons, that are necessary ingredients in a technicolor model, we need to include the fluctuations of the worldvolume (techniflavour) field strength $F_{ab}$ of the D7 branes, with $a,b = 0,1,...,7$. In other words, we have to consider the DBI action
\be \label{SDBI}
S_{D7} = -T \int d^4 x \,d \rho \,d \Omega_3 \,e^{- \phi} \sqrt{- \det (g_{ab} + 2 \pi \alpha' F_{ab})} \,\, ,
\ee
where $g_{ab}$ is the metric induced on the D7 worldvolume, $\Omega_3$ is the 3-sphere wrapped by the D7 branes 
and $\phi$ is the dilaton. To leading, quadratic, order in $F_{ab}$ the above action becomes:
\be \label{YM}
S_{D7} = -T (2 \pi \alpha')^2 \int d^4 x \,d \rho \,d \Omega_3 \,e^{-\phi} \sqrt{- \det (g_8)} \,{\rm Tr} g^{ab} g^{cd} F_{ac} F_{bd} \, . 
\ee
Using the embedding solution (\ref{VPrho0}), together with $\theta = \pi /2$, and integrating over $\Omega_3$, we obtain:
\be \label{GFAction}
S_{D7} = - \frac{\kappa}{4} \int d^4 x \, d \rho \,\left[ a(\rho) F_{\mu \nu} F^{\mu \nu} + 2 b(\rho) F_{\mu \rho} F^{\mu}{}_{\rho} \right] \, ,
\ee
where $\mu, \nu = 0,1,2,3$ and $\kappa = \frac{T (2 \pi \alpha')^2 V_3}{g_s}$ with $V_3$ being the volume of the compact cycle wrapped by the D7 probe brane. Furthermore, the functions $a(\rho)$ and $b(\rho)$ are:
\bea \label{ab}
a (\rho) &=& \frac{1}{A^3} \left[ f(\rho) + g(\rho) \varphi_{\rho}^2 \right]^{1/2} \nn \\
b (\rho) &=& \frac{1}{A^2} \left[ f(\rho) + g(\rho) \varphi_{\rho}^2 \right]^{1/2} g^{\rho \rho} \,\, ,
\eea
where the inverse world-volume metric component $g^{\rho \rho}$ is given by
\be \label{grr}
g^{\rho \rho} = \frac{A^7 c^3 P_1^2 P_1' \coth^2 (2 \rho)}{128 \left( f+g \varphi_{\rho}^2 \right)} = \frac{2}{Ac \left[ P_1' + \frac{1}{2} P_1 \tanh (2 \rho) \,\varphi_{\rho}^2  \right]} \,\, .
\ee
Finally, let us determine explicitly the coefficient $\kappa$ in (\ref{GFAction}). Since $V_3$ is the surface "area" of a unit 3-sphere, it is given by $V_3 = 2\pi^2$. Then, recalling that we work in units of $l_s^2 = \alpha' = 1 = g_s$\,, we have:
\be \label{kappa}
\kappa = \frac{1}{(2 \pi)^7 l_s^8} (2 \pi \alpha')^2 2 \pi^2 = \frac{1}{2} \frac{1}{(2 \pi)^3} \,\, .
\ee

To conclude the description of the gravity dual set-up, let us also recall that the vector and axial-vector modes we are interested in arise from the following expansion of the techniflavour gauge potential $A_{\mu} (x,\rho)$:
\be \label{decomp}
A_{\mu} (q, \rho) = {\cal V}_{\mu} (q) \psi_V^0 (q^2, \rho) + {\cal A}_{\mu} (q) \psi_A^0 (q^2, \rho) + \sum_{n} \left( V_{\mu}^n (q) \psi_{V_n} (\rho) + A_{\mu}^n (q) \psi_{A_n} (\rho) \right) \, ,
\ee
where we have Fourier transformed in $x^{\mu}$. Furthermore, the modes in the sum over $n$ are the normalizable ones, while $\psi_{V}^0$ and $\psi_A^0$ are the non-normalizable modes that correspond to sources for the vector and axial-vector field theory currents, respectively.\footnote{Note that, just like \cite{HolTech},  in (\ref{decomp}) we neglect an additional term proportional to the derivative of the technipion field, as that is not important for our purposes. Such a term arises from the gauge transformation that sets $A_{\rho} = 0$. It plays an essential role in obtaining the technipion's kinetic term and its interactions with other fields in the four-dimensional Lagrangian, just as in \cite{SS2}.} An important point is that, as in the Sakai-Sugimoto model \cite{SS}, the vector modes are those that are symmetric w.r.t. reflection (on the probe brane embedding) around $\rho_0$, whereas the axial modes are those that are antisymmetric w.r.t. this operation. For more details on the expansion modes see \cite{LA}. Here we will only write down, for future use, the normalization condition
\be \label{norm}
\kappa \int_{{\rm D}7+\overline{{\rm D}7}} d\rho \,\, a(\rho) \, \psi_n \psi_m = \delta_{nm} \,\, ,
\ee
that is needed (both for V and for A modes) in order for the four-dimensional action to be canonically normalized, as well as the equations of motion 
\be \label{normFE}
\pd_{\rho} \,[\,b(\rho) \,\pd_{\rho} \psi_n (\rho)\,] = - m_n^2 \,a(\rho) \,\psi_n (\rho)
\ee
and
\be \label{NNMode}
\pd_{\rho} \,[\,b(\rho) \,\pd_{\rho} \psi_{V,A}^0 (q^2, \rho)\,] = - q^2 \,a(\rho) \,\psi_{V,A}^0 (q^2, \rho) \,\, ,
\ee
that follow from the action (\ref{GFAction}) with (\ref{decomp}) substituted.

\subsection{Condition for regular axial-vector modes}

Let us now take a closer look at the equation
\be\label{Eqrho}
b (\rho) \,\psi'' (E^2, \rho) + b' (\rho) \,\psi' (E^2, \rho) + E^2 a(\rho) \,\psi(E^2, \rho) = 0 \, ,
\ee
where $' \equiv \pd_{\rho}$ and $E$ denotes any of $ m_{V_n}, m_{A_n}, q $. We will show that there is an interesting condition, that has to be satisfied in order for this equation to have regular axial-vector solutions $\psi_{A_n}$. In order to do that, it will be convenient to change variables from $\rho$ to a worldvolume coordinate $z$ that distinguishes between the D7 and the $\overline{{\rm D}7}$ branches. Furthermore, since vector (axial-vector) modes are defined via their symmetry (anti-symmetry) with respect to reflection around the point $\rho=\rho_0$ on the D7-$\overline{{\rm D}7}$ embedding, the coordinate $z$ should be such that $\rho=\rho_0$ corresponds to $z=0$. A suitable choice, discussed in \cite{LA}, is $z^2 = \rho^2 - \rho_0^2$, where $z>0$ ($z<0$) gives the D7 ($\overline{{\rm D}7}$) branch. Another choice, that will turn out to be natural in our context, will be introduced in Section \ref{DiscrSt}. For any change of variable $\rho \rightarrow \rho(z)$ though, one finds that the equation of interest becomes:
\be \label{Eqz}
\hat{b} (z) \,\psi'' (E^2,z) + \hat{b}' (z) \,\psi' (E^2,z) + E^2 \hat{a}(z) \,\psi (E^2,z) = 0 \, ,
\ee
where now $' \equiv \pd_z$ and \cite{LA}:
\be \label{abz}
\hat{a}(z) = a(\rho) \,\pd_z \rho \qquad , \qquad \hat{b}(z) = \frac{b(\rho)}{\pd_z \rho}  \, .
\ee
Note that, clearly, physical observables should not depend on the choice of $z$. This is ensured by the relations (\ref{abz}), as will become evident in Section \ref{Sexact}. On conceptual level, the independence of physical results on the concrete function $\rho (z)$ is due to the fact that different choices of $z$ differ only by a worldvolume coordinate transformation.

Now, we are looking for two types of solutions of (\ref{Eqz}): vector modes, which are defined by being symmetric under $z \rightarrow -z$ and thus satisfy $\psi_V'(0) = 0$, and axial-vector modes, that are antisymmetric under $z \rightarrow - z$ and so $\psi_A (0) = 0$. As it turns out, there is a nontrivial condition for the existence of regular antisymmetric solutions. This is due to the different behaviours at $z=0$ of the functions of interest when $g(\rho_0) \neq e^{4 \rho_0}$ and when $g(\rho_0) = e^{4 \rho_0}$. To see this, let us consider in more detail the small $z$ expansions of the various relevant expressions. We will only need that $\pd_z \rho \sim z$ at small $z$ for any change of variables of the form $z^2 = \tilde{f} (\rho - \rho_0)$. It is easy to realize that $f(\rho)$ and $g(\rho)$ in (\ref{fg}) are regular at $\rho\approx \rho_0$, or equivalently at $z\approx 0$. On the other hand, $\varphi_{\rho}^2$ in (\ref{VPrho0}) behaves as:
\be
\varphi_{\rho}^2 \sim  \begin{cases}const + {\cal O} (z^2)\,,& \mbox{when $g(\rho_0) \neq e^{4 \rho_0}$}\\\frac{1}{z^2} + ... \,, & \mbox{when $g(\rho_0) = e^{4 \rho_0}$\,.}\end{cases}
\ee
Using this in (\ref{ab}) and (\ref{grr}), one can extract the small $z$ behaviour of $\hat{a} (z)$ and $\hat{b} (z)$.
%Then we obtain from (\ref{grr}) that
%\be
%g^{\rho\rho} \sim  \begin{cases}\mbox{const}& \mbox{if $\cal C$ is not satisfied}\\z^2, & \mbox{if $\cal C$ is satisfied}\end{cases}
%\ee
As a result, one finds that there are always even (vector) solutions of (\ref{Eqz}), that are regular at $z=0$. However, regular odd (axial-vector) solutions exist only for $g(\rho_0) = e^{4 \rho_0}$. When this condition is not satisfied, the odd modes develop a logaritmic singularity at $z=0$.

Given the importance of the condition $g(\rho_0) = e^{4 \rho_0}$ for the existence of a technicolor model, let us write it down more explicitly:
\be \label{Rel}
\frac{3}{256} \!\left[ 8 (1-\beta^{2/3})^{3/2} \!+ 4 \beta (e^{4 \rho_0} - e^{-4\rho_0} - 8 \rho_0) \right]^{1/3} \!(e^{4 \rho_0} - e^{-4 \rho_0}) = 2 c^2 \beta e^{4 \rho_0} \, ,
\ee
where for convenience we have introduced the notation $\beta \equiv \sin^3 \alpha$ and used that $\cos^3 \alpha = (1 - \beta^{2/3})^{3/2}$. This relation allows us to solve for one of the three parameters $\beta$, $c$ and $\rho_0$ in terms of the other two. So we end up with one less free parameter than originally expected. Let us also comment on the consistency of this condition with the necessary  parameter ranges, more precisely:
\be \label{Cond}
c >\!\!> 1 \qquad {\rm and} \qquad \beta <\!\!< 1 \, ,
\ee
as recalled under (\ref{simpM}). It is easy to see, keeping $\rho_0$ fixed for simplicity, that if one increases the value of $c$, then (\ref{Rel}) leads to decreasing the value of $\beta$. So there are infinitely many solutions consistent with (\ref{Cond}). This is even more evident upon realizing that the constraint (\ref{Rel}) simplifies significantly at leading order in small $\beta$, due to the fact that $\rho_0$ is always at least of order 1. Namely, we have:
\be \label{SimpConstr}
\sqrt{3} \approx 16 c \sqrt{\beta} \, .
\ee

\section{S-parameter: exact result} \label{Sexact}
\setcounter{equation}{0}

In this section we will compute the electroweak $S$ parameter using the exact formula of \cite{HolTech,LA}, that follows from the gauge/gravity duality statement that the generating functional of the field theory correlators is the dual gravitational action. This relation allows one to extract a compact answer in terms of the non-normalizable modes introduced in the previous section. 
To find those modes, we will use a different approach compared to \cite{LA}, which will not leave any undetermined constants at leading order in small $\beta$. 

\subsection{Preliminaries} 

The S-parameter is defined as \cite{PT}:
\be \label{Spar}
S = - 4 \pi \frac{d}{d q^2} \left( \Pi_V - \Pi_A \right)\bigg|_{q^2=0} \, ,
\ee
where $\Pi_V$ and $\Pi_A$ are the two-point correlators for the vector and axial-vector currents respectively. 
In \cite{LA}, it was shown that:
\bea \label{PiVA}
\Pi_V (q^2) &=& 2 \kappa \left[ b(\rho) \,\psi_V^0 (q^2, \rho) \,\pd_{\rho} \psi_V^0 (q^2, \rho) \right]_{\rho= \infty} \nn \\
\Pi_A (q^2) &=& 2 \kappa \left[ b(\rho) \,\psi_A^0 (q^2, \rho) \,\pd_{\rho} \psi_A^0 (q^2, \rho) \right]_{\rho=\infty} \, ,
\eea
where the constant $\kappa$ is given in (\ref{kappa}), $\psi^0_{V,A}$ are the solutions of (\ref{NNMode}) and, finally, $b(\rho)$ is the same as in (\ref{ab}). Recalling that the technipion decay constant is given by $F_{\pi}^2 = \Pi_A (0)$, we can see from (\ref{PiVA}) that:
\be \label{Fpi2}
F_{\pi}^2 = 2 \kappa \left[ b(\rho) \,\psi_A^0 (0, \rho) \,\pd_z \psi_A^0 (0, \rho) \right]_{\rho=\infty} \, .
\ee

However, applying the above formulae to the present case requires a slight modification (something that was not realized in \cite{LA}). The reason is that the background of interest has a walking region and a UV region and the transition between these two regions is related to a change of spectrum in the field theory. That was not the case for the duals of regular (i.e. QCD-like) technicolor, for which there was just a single region and so the integration in $\rho$ was going all the way to $\infty$. In the present case however, such an integration would not encode correctly the relevant physics, since $S$ should be determined by the modes in the walking region; the modes in the UV region, on the other hand, should somehow be related to extended technicolor.\footnote{Whether (or not) the background of \cite{NPP}, as well as the necessary techniflavour probe embeddings, require some modification in order to properly account for this extended field theory is an interesting open question, that we hope to address in the future.} So in the following we will introduce a physical cut-off $\rho_{\Lambda}$, which is the upper end of the walking region. Hence, instead of (\ref{PiVA}) and (\ref{Fpi2}), we should compute the S-parameter and $F_{\pi}^2$ from:
\be \label{Scutoff0}
S = - 8 \pi \kappa \,\, b(\rho) \,\frac{d}{d q^2} \left[ \psi_V^0 (q^2, \rho) \,\pd_{\rho} \psi_V^0 (q^2, \rho) - \psi_A^0 (q^2, \rho) \,\pd_{\rho} \psi_A^0 (q^2, \rho) \right]\bigg|_{\rho=\rho_{\Lambda},\,q^2=0} 
\ee 
and
\be
F_{\pi}^2 = 2 \kappa \left[ b(\rho) \,\psi_A^0 (0, \rho) \,\pd_z \psi_A^0 (0, \rho) \right]_{\rho=\rho_{\Lambda}} \, ,
\ee
respectively. Note that this is similar to \cite{APTNPR}, where a physical cut-off was introduced in the computation of glueball and meson spectra in (deformations of) the Maldacena-Nunez background, that is dual to ${\cal N}=1$ SYM.\footnote{As an interesting aside, let us also mention that the introduction of a cut-off in the QCD-like technicolor model, based on the Sakai-Sugimoto brane set-up, played an important role in reducing the value of the $S$ parameter in that model compared to standard expectations \cite{AB}.}

In view of the above, in order to satisfy the boundary condition $\psi^0 (\rho_{\Lambda}) = 1$ we have to take:
\be
\psi^0_{A,V} (\rho) \, \rightarrow \, \frac{\psi^0_{A,V} (\rho)}{ \psi^0_{A,V} (\rho_{\Lambda})} \, .
\ee
To incorporate this normalization automatically, we write $\Pi_{A,V}$ as:
\bea \label{PiVA2}
\Pi_V (q^2) &=& 2 \kappa \left. b(\rho)\,\pd_\rho \log[\psi_V (q^2, \rho) ]\right|_{\rho= \rho_{\Lambda}} \nn \\
\Pi_A (q^2) &=& 2 \kappa \left.b(\rho)\,\pd_\rho \log[\psi_A (q^2, \rho) ]\right|_{\rho= \rho_{\Lambda}}.
\eea
From now on, we will use these expressions to evaluate (\ref{Spar}). Before turning to the computation of the S-parameter though, let us first make a few comments about the cutoff $\rho_{\Lambda}$.

In \cite{NPP} it was shown that the upper end of the walking region is roughly given by (\ref{cutoffNPP}). This estimate is obtained by investigating (\ref{g2TC}), i.e. the behaviour of the technicolor gauge coupling $g^2$ as a function of the radial variable $\rho$. One can see that the walking region corresponds to the sum in the denominator being dominated by $\cos^3 \alpha$. Hence, a rough estimate for $\rho_{\Lambda}$, the upper end of the walking region, follows from the condition that $\sin^3 \alpha \left( \sinh (4 \rho) - 4 \rho \right)$ becomes comparable with $\cos^3 \alpha$. In other words: $\frac{1}{2} e^{4 \rho_{\Lambda}} \approx \cot^3 \alpha$, since $\rho$ is always at least of order 1 or greater; this is the condition (\ref{cutoffNPP}). However, this cutoff is, in fact, in the middle of the smooth transition between the walking and the UV regions. A more careful estimate, that removes this transition region, is provided by
\be \label{upperend}
\frac{1}{2} e^{4\rho_{\Lambda}} = 0.02 \times \cot^3 \alpha \,\, .
\ee
Note also that we can extract an upper bound, $\beta_u$, on the parameter $\beta = \sin^3 \alpha$. Namely, since the walking region is in the interval $\rho \in (1, \rho_{\Lambda})$ and greater values of $\beta$ correspond to smaller $\rho_{\Lambda}$, we can obtain the greates value $\beta_u$, above which there is no walking, by taking $\rho_{\Lambda}=1$ in (\ref{upperend}). This gives:
\be
\beta_u = 7.2 \times 10^{-4} \, .
\ee 
In other words, to have a walking region at all, we need $\beta < \beta_u$. Clearly, the less stringent cutoff arising from $\frac{1}{2} e^{4 \rho_{\Lambda}} \approx \cot^3 \alpha$ gives higher upper bound on $\beta$, namely $\beta_u = 3 \times 10^{-2}$.

\subsection{Computation}

Now we will compute the solutions of (\ref{NNMode}) order by order in small $q^2$. We will use a different approach compared to \cite{LA}, which will allow us (due to $\beta <\!\!< 1$) to calculate analytically the constants that remain undetermined by the method used there. Namely, integrating (\ref{NNMode}) twice, we obtain the following integral equation:
\be \label{inteq}
\psi_{V,A}^0(\rho, q^2)=c_1
+c_2 \!\int_{\rho_0}^\rho \frac{1}{b(\rho_1)} d\rho_1 - q^2 \! \int_{\rho_0}^\rho  \frac{1}{b(\rho_1)}\left(\int_{\rho_0}^{\rho_1} \!a(\rho_2) \,\psi_{V,A}^0(\rho_2, q^2) \,d\rho_2 \!\right) \!d\rho_1 \, ,
\ee
where $c_{1,2}$ are arbitrary constants. This can be solved order by order in small $q^2$ by iteration. In particular, in order to find the first order solution, we need to substitute the zeroth order $\psi^0_{A,V}$ in the last term. Note also that, as is easy to realize, under reflection $z \rightarrow - z$ of the worldvolume variable $z$ the first term in (\ref{inteq}) is even, while the second one is odd. The third term, on the other hand, is even (odd) when $\psi^0_{A,V}$ is even (odd). So, to leading order in small $q^2$, we have that:
\bea\label{inteqVA}
\psi_V^0(\rho,q^2)&=&1-q^2\int_{\rho_0}^\rho  \frac{1}{b(\rho_1)}\left(\int_{\rho_0}^{\rho_1} a(\rho_2)\,d\rho_2\right)d\rho_1 + {\cal O} (q^4) \, ,\nn\\
\psi_A^0(\rho,q^2)&=&\!\int_{\rho_0}^\rho \frac{1}{b(\rho_1)}d\rho_1 \left( 1 -q^2 \int_{\rho_0}^{\rho_1} a(\rho_2)\int_{\rho_0}^{\rho_2} \frac{1}{b(\rho_3)} d\rho_3\,d\rho_2 \right) \!+ {\cal O} (q^4) \, ,
\eea
where we have taken $c_{1,2} = 1$ for easier comparison with \cite{LA} as will become clear shortly. Using that in the walking region the expressions (\ref{ab}) acquire the form
\bea\label{absimp}
a(\rho)&=&  \frac{1}{24}3^{1/4}c_0^{5/4} \beta^{1/8}e^{4\rho}\sqrt{\frac{1}{1-e^{4(\rho_0-\rho)}}} \, , \nn\\
b(\rho)&=&\frac{1}{8}3^{1/4}c_0^{1/4}\beta^{-3/8}\sqrt{1-e^{4(\rho_0-\rho)}} \, ,
\eea
where we have introduced the notation $c_0 \equiv c \sqrt{\beta}$, we can easily compute the integrals in (\ref{inteqVA}). As a result, we obtain to leading order in small $\beta$:
\be\label{S1}
S=\frac{\kappa\, c_0^{5/4}\pi \,\beta^{1/8} \,e^{4 \rho_{\Lambda}}(L-1)}{2\times 3^{3/4} L^2} 
\ee
and
\be\label{fpi}
F_\pi^2=\frac{3^{1/4}\kappa\, c_0^{1/4}M_{KK}^2}{\beta^{3/8}L} \, ,
\ee
where $L = 4 (\rho_{\Lambda} - \rho_0) + \log 4$ and so $L-1 > 0$. Note that, due to (\ref{SimpConstr}), we have to leading order: $c_0 \approx \sqrt{3} / 16$. 

Let us make a couple of important remarks about (\ref{S1}). First, it clearly implies that $S>0$ in our model. And second, because $\rho_{\Lambda}$ is a function of $\beta$ (recall (\ref{upperend})), the overall dependence of $S$ on $\beta$ is $S \sim \beta^{-7/8}/\log(\beta)$. As a result, decreasing $\beta$ means increasing $S$, which is rather unexpected. At this point, it is not at all clear whether this is a generic prediction of holographic walking technicolor or a strange peculiarity of the model we are considering here. We hope to come back to this issue in the future. In any case, we will see in the next section that the answer (\ref{S1}) is in perfect agreement with the result of an independent computation, obtained from summing the contributions of the discrete vector and axial-vector modes.

Finally, for comparison with \cite{LA}, note that solving the differential equation (\ref{Eqrho}) iteratively in small $q^2$ one finds:
\be \label{WalkSol}
\psi_{V,A}^0 (\rho) = 1 + C_1^{V,A} q^2 + (C_2^{V,A} + C_3^{V,A} q^2) \,\rho - \frac{q^2 Q}{32} \,e^{4 \rho} \,(2-C_2^{V,A} +2 C_2^{V,A} \rho) + {\cal O} (q^4) \,\, , 
\ee
where $Q \equiv \frac{a(\rho)}{b(\rho)} e^{-4 \rho}$. One can extract the coefficients $C_{1,2,3}^{V,A}$\,\,, whose analogues in \cite{LA} remained undetermined, by matching (\ref{WalkSol}) with (\ref{inteqVA}).\footnote{For completeness, let us record here the answer to leading order in small $\beta$:
\bea 
&&C_1^V = \frac{c_0 \beta^{1/2} e^{4 \rho_0} (1-4\rho_0)}{48} \,\, , \qquad C_2^V = 0 \,\, , \qquad C_3^V = \frac{c_0 \beta^{1/2} e^{4 \rho_0}}{12} \,\, , \nn \\
&&C_1^A = \frac{c_0 \beta^{1/2} e^{4 \rho_0} (1-2 \rho_0 )}{96 \rho_0} \,\, , \qquad C_2^A = -\frac{1}{\rho_0} \,\, , \qquad C_3^A = \frac{c_0 \beta^{1/2} e^{4 \rho_0}}{48 \rho_0} \,\, .
\eea} 
Then, substituting (\ref{WalkSol}) in
%To illustrate this, let us consider the simplified expressions:
%\be
%a(\rho) \approx \frac{c_0^{5/4}}{8\times 3^{3/4}} \,\beta^{1/8} \,e^{4\rho} \qquad {\rm and} \qquad b(\rho) \approx \frac{3^{1/4} c_0^{1/4}}{8} \,\frac{1}{\beta^{3/8}} \,\,\, ,
%\ee
%valid for $\rho_0 <\!\!< \rho$.
\be
S = - 8 \pi \kappa \left[ b(\rho ) \,\frac{\pd}{\pd q^2} \left( \pd_{\rho} \log \psi_V^0 - \pd_{\rho} \log \psi_A^0 \right) \right]_{\rho = \rho_{\Lambda}, \,q^2=0} \, ,
\ee
we again find (\ref{S1}).
%\be \label{Scutoff2}
%S = - \frac{\pi \kappa \,3^{1/4} c_0^{5/4} \beta^{1/8} \left[ 8 e^{4\rho_0} (\rho_0 - \rho_{\Lambda})^2 + 4 e^{4\rho_{\Lambda}} (\rho_0 - \rho_{\Lambda}) - e^{4\rho_0} + e^{4 \rho_{\Lambda}}\right]}{96 (\rho_0 - \rho_{\Lambda})^2} \, ,
%\ee

\section{Discrete states} \label{DiscrSt}
\setcounter{equation}{0}

In this section we will investigate the spectrum of vector and axial-vector mesons. To begin with, we will rewrite the relevant equations of motion in a Schrodinger form. The change of variables, that enables that, will provide us with a natural worldvolume radial coordinate $z$ that runs over the whole D7-$\overline{{\rm D}7}$ embedding. Then we will calculate the masses and decay constants of the V and A modes. Finally, we will compute the S parameter by summing the discrete state contributions and will find complete agreement with the exact result of the previous section, in contrast with the situation in \cite{HolTech}. 

\subsection{Schrodinger form of field equation}

In order to transform the field equation (\ref{Eqrho}) into Schrodinger form, let us introduce the following variable (very much like the tortoise variable in General Relativity):
\be\label{zdef}
z=\pm \int_{\rho_0}^\rho \left(\frac{a(\rho')}{b(\rho')}\right)^{1/2}d\rho',
\ee
where the "$+$" sign corresponds to the D7 branch and the "$-$" sign to the $\overline{{\rm D}7}$ one. Substituting (\ref{absimp}) into (\ref{zdef}), we find:
\be\label{zexp}
z=\pm \sqrt{\frac{c_0}{12}}\,\beta^{1/4}\sqrt{e^{4\rho}-e^{4\rho_0}} \,.
\ee
Note that near $\rho=\rho_0$ the right-hand side of (\ref{zexp}) behaves as $(\rho-\rho_0)^{1/2}$. This implies that $z^2\sim \rho-\rho_0$ for $\rho\approx\rho_0$, similarly to the choice of variable $z$ considered in \cite{LA}. 
Let us also introduce a function $\phi (z)$ defined via:
\be\label{newwave}
\psi(\rho)=\frac{1}{[a(\rho)b(\rho)]^{1/4}}\phi(z) \,,
\ee
where $[a(\rho)b(\rho)]^{1/4}$ is viewed as a function of $z$ via (\ref{zexp}) and, for convenience, we have dropped the index $n$.
%for brevity, we have dropped the argument $E^2=m^2$. 
With these definitions, one can show that (\ref{normFE}) implies the following differential equation for $\phi(z)$:
\be\label{Sch1}
-\phi''(z)+V(z)\phi(z)=m^2\,\phi(z) \, ,
\ee
where the Schrodinger potential is given by
\be\label{V2}
V(z)=-\frac{1}{4}\frac{z^2-2 \lambda}{(z^2+\lambda)^2}
\ee
with
\be
\lambda=\frac{1}{12}e^{4\rho_0}\beta^{1/2}c_0 \, .
\ee
Since $\rho \leq \rho_{\Lambda}$\,, clearly we have that $-z_{\Lambda} \leq z \leq z_{\Lambda}$\,, where
\be\label{zmax}
z_\Lambda=\sqrt{\frac{c_0}{12}}\,\beta^{1/4}\sqrt{e^{4\rho_\Lambda}-e^{4\rho_0}}\simeq\sqrt{\frac{c_0}{12}}\,\beta^{1/4}e^{2\rho_\Lambda} \, .
\ee
Note also that (\ref{upperend}) implies that $e^{2 \rho_{\Lambda}} \sim \beta^{-1/2}$ and hence $z_{\Lambda} = {\cal O} (\beta^{-1/4}) >\!\!> 1$.

In the following subsections we will study the Schrodinger equation (\ref{Sch1}) in order to find the mass spectra and decay constants of the vector and axial-vector bosons in our model. 
Before proceeding, let us mention that the above method of transforming equation (\ref{Eqrho}) to Schrodinger form is equally applicable to the models studied in \cite{SS} and \cite{HolTech}. In their case, one finds that the tortoise variable $z$ has a finite range, i.e. $z \leq z_{\star}$ for some $z_{\star}$, and the Schrodinger potential behaves as $V(z)\to\infty$ for $z\to z_{\star}$. As a result, there is an infinite spectrum of vector and axial-vector mesons.

%XXXXXXXXXXXXXXXXXXXXXXXXXXXXXXXXXX
%
%The potential, when we transform it back to variable $\rho$ takes the form $W(\rho)=V(z)$
%\be\label{Wrho}
%W(\rho)=\frac{2ab(a'b'+2ab'')+b^2[4aa''-5(a')^2]-a^2(b')^2}{16a^3b},
%\ee
%where the argument, $\rho$, has been dropped for simplicity and where the derivatives are taken with respect to $\rho$.

\subsection{Mass spectrum} \label{MassSpec}

Now we will compute the mass spectrum of the solutions of equation (\ref{Sch1}) to leading order in small $\beta$. To do that, we will split the range of variation of $z$ (on the positive branch) into two subintervals, in which exact solutions can be found, and then we will match those two solutions in a region, where they behave in the same way. Namely, we take the two subintervals to be $(0,z_{\bullet})$ and $(z_{\bullet},z_{\Lambda})$, where $z_{\bullet}=N\sqrt{\lambda}$ with $N$ being a number satisfying $1<\!\!< N <\!\!< \beta^{-1/4}$. 

In the first subinterval, we will keep the form of $V(z)$ in (\ref{V2}) unchanged but will omit the right-hand side of (\ref{Sch1}). This will be justified a posteriori, since we will see that $V(z)={\cal O}(\beta^{-1/2})>\!\!>m^2={\cal O}(\beta^{1/2})$ once we compute the mass spectrum. With these assumptions, the solution $\phi_1 (z)$ of (\ref{Sch1}) in this interval is: 
%satisfying Dirichlet  boundary conditions for vector and Neumann boundary conditions for axial vector bosons at $z=0$ is
\be\label{psi1}
\phi_1(z)\simeq \begin{cases}(z^2+\lambda)^{1/4},& \mbox{\,\,for symmetric (vector) modes\,,}\\(z^2+\lambda)^{1/4}\log\left(\frac{z+\sqrt{z^2+\lambda}}{-z+\sqrt{z^2+\lambda}}\right), & \mbox{\,\,for antisymmetric (axial) modes\,.}
\end{cases}
\ee

In the second interval, $z_{\bullet}<z<z_{\Lambda}$\,, we have that $z^2>\!\!>\lambda$ and so we can approximate the potential by $V(z)\simeq-1/(4z^2)$. Then the general solution $\phi_2 (z)$ of (\ref{Sch1}), satisfying the boundary condition $\phi_2 (z_{\Lambda}) = 0$, is:
\be\label{phi2}
\phi_2(z)\simeq\sqrt{z}[J_0(m\,z)-R\,Y_0(m\,z)]\,,
\ee
where $R=J_0(m\,z_{\Lambda})/Y_0(m\,z_{\Lambda})$.

Now, let us match the logarithmic derivatives of $\phi_1 (z)$ and $\phi_2 (z)$ at $z=z_{\bullet}$, in order to obtain the discrete spectrum of vector and axial-vector modes. We start by computing  
%{\color{blue} The spectrum of bound states is obtained from matching the logarithmic derivatives} of $\phi_1(z)$ and $\phi_2(z)$  at $z = z_{\bullet}$. 
%We obtain 
from (\ref{psi1}), to leading order in small $\lambda/z_{\bullet}^2$:
\be\label{psi0}
2z_{\bullet}\frac{\phi_1'(z_{\bullet})}{\phi_1(z_{\bullet})} \simeq \begin{cases}1\,,& \mbox{\,\,for vector bosons\,,}\\1+\frac{2}{\log(2 z_{\bullet} / \sqrt{\lambda})}\,, & \mbox{\,\,for axial-vector bosons\,.}
\end{cases}
\ee
Similarly, from (\ref{phi2}) we find to leading order in small $\beta$:
\be\label{cond}
2z_{\bullet}\frac{\phi_2'(z_{\bullet})}{\phi_2(z_{\bullet})}=1-2\mu\frac{\,J_1(\mu)-R\,Y_1(\mu)}{\,J_0(\mu)-R\,Y_0(\mu)}\simeq1
+\frac{2\,R}{R[\log(m/2)+\,\log(z_\bullet)+\gamma]-\pi\,/\,2}\,,\ee
where $\mu \equiv m\,z_{\bullet}$ and $\gamma$ is Euler's constant; on the right-hand side of (\ref{cond}) we substituted the leading small-argument asymptotics of the Bessel functions since $\mu <\!\!< 1$. Equating (\ref{psi0}) and (\ref{cond}), we obtain the following quantization conditions:
\be
R=0 \quad \Rightarrow \quad J_0 (m z_{\Lambda}) = 0 \,, \qquad \,\,{\rm for \,\,vector \,\,bosons} 
\ee
and 
\be
R=\frac{\pi}{2}\left[\gamma +\log\left( \frac{m\sqrt{\lambda}}{4}\right)\right]^{-1}\,, \qquad
{\rm for \,\,axial\!-\!vector \,\,bosons} \, . 
\ee
Thus, the mass spectrum of vector bosons is given by $m_n^V z_{\Lambda}=r_n^J$ with $r_n^J$ being the $n^{\rm th}$ root of $J_0$. In other words, restoring dimensions, we have:
\be \label{Vmass}
m_n^{V}=\frac{2\sqrt{3}}{\sqrt{c_0}}\beta^{-1/4}e^{-2\rho_\Lambda}r_n^{J}M_{KK} \, ,
\ee 
where $M_{KK}$ is a mass scale that is determined by setting $F_{\pi}^2 = (246 \,{\rm GeV})^2$ 
in (\ref{fpi}).
%\footnote{ See comments concerning the appropriateness of this choice in the next section.} 
Similarly, the mass spectrum of the axial-vector bosons is:
\be \label{Amass}
m_n^{A}=\frac{2\sqrt{3}}{\sqrt{c_0}}\beta^{-1/4}e^{-2\rho_\Lambda}\mu_n M_{KK} \, ,
\ee
where $\mu_n$ is the $n^{\rm th}$ root of
\be\label{axialmass}
J_0( \mu_n)\,\frac{2}{\pi}\left[\gamma +\log\left( \frac{\mu_n\sqrt{\lambda}}{4\,z_\Lambda}\right)\right] \!= \,Y_0(\mu_n) \, .
\ee
Note that since $e^{2 \rho_{\Lambda}} \sim \beta^{-1/2}$, all of the masses $m_n^{V,A}$ are of order $\beta^{1/4}$.

\begin{figure}[t]
\begin{center}
\includegraphics[width=4in]{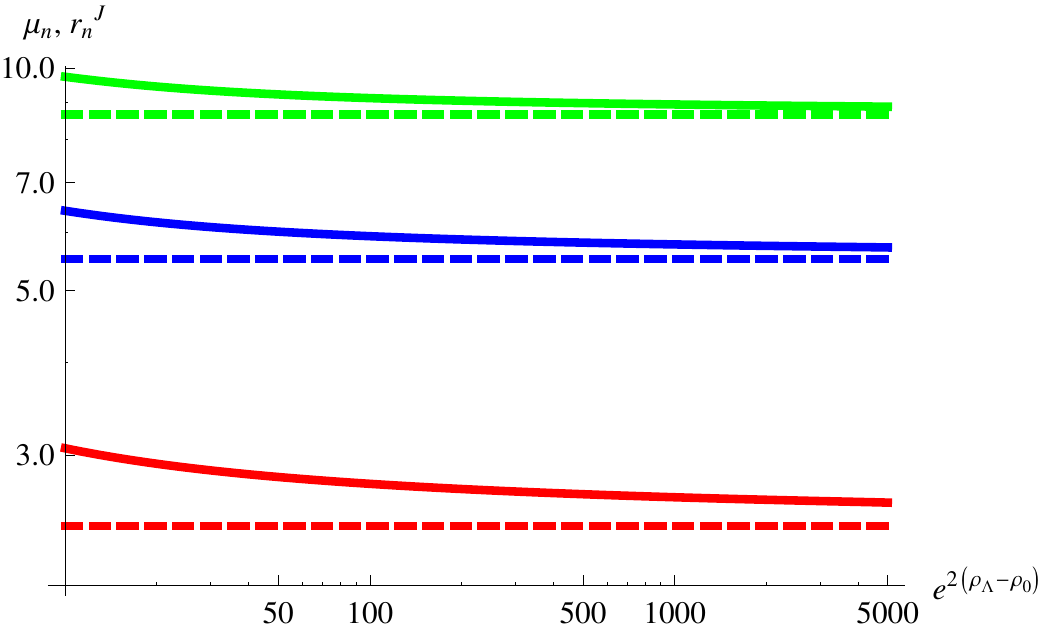}
\caption{$\mu_n$ (solid lines) and $r_n^J$ (dashed lines) for the three lightest vector and axial-vector mesons, as a function of $e^{2(\rho_\Lambda-\rho_0)}=z_\Lambda\,/\, \sqrt{\lambda}$\,. 
%Solid lines represent $\mu_n$ (i.e., axial-vectors), while dashed ones represent $r_n^J$ (i.e., vectors).
}
\label{plotmass}
\end{center}
\end{figure}

Although (\ref{axialmass}) cannot be solved analytically, it is easy to evaluate  its solutions numerically at arbitrary $n$ and  $z_\Lambda/\sqrt{\lambda}=e^{2(\rho_\Lambda-\rho_0)}$. Through the latter quantity, $\mu_n$ is a function of the cutoff $\rho_{\Lambda}$. We argued in Sec. \ref{Sexact} that the cutoff in $\rho$ should be determined by (\ref{upperend}), instead of the less precise estimate (\ref{cutoffNPP}) given in \cite{NPP}. Despite that, for more generality, let us now address how the solutions $\mu_n$ depend on the value of $\rho_{\Lambda}$. The choice $e^{4 \rho_{\Lambda}}\beta\,/\,2 = 0.02$ is the same as in (\ref{upperend}), whereas $e^{4 \rho_{\Lambda}}\beta\,/\,2 = 1$ would correspond to $\rho_{\Lambda} = \rho_*$.  So in order to illustrate the dependence  on the choice of cutoff, as well as how axial-vector boson masses compare to vector boson masses, we plot $\mu_n$ and  $ r_n^J$ for $n=1,2,3$ on Fig.\ref{plotmass} as functions of $e^{2(\rho_\Lambda-\rho_0)}$.
%\begin{figure}[htbp]
Note that at low orders one has $R<\!\!<1$. Thus the axial wave function is dominated by the term $\sqrt{z}\,J_0(m\,z)$ and the axial-vector masses (\ref{Amass}) are close to the vector ones (\ref{Vmass}). Furthermore, the difference $m_n^A - m_n^V$ decreases as $\beta$ gets smaller. Nevertheless, the vector bosons remain always lighter than the axial ones (at every order), just like in QCD.

\subsection{Decay constants and S-parameter}

In this subsection, we will compute the decay constants $g_{V_n},g_{A_n}$ for the vector and axial-vector modes. This will allow us to obtain another estimate for the S-parameter, by utilizing the formula \cite{PT}:
\be\label{sumrule}
S=4\pi\sum_{n=1}^\infty\left(\frac{g_{V_n}^2}{m_{V_n}^4}-\frac{g_{A_n}^2}{m_{A_n}^4}\right)\,.
\ee
Unlike \cite{HolTech}, we will find that the result from (\ref{sumrule}) agrees completely with the answer obtained from the exact method of Section \ref{Sexact}.

In \cite{LA}, it was shown that
\bea\label{coupling2}
g_{V_n}&=&- \kappa\, \int_{{\rm D7 + \overline{D7}}}\psi_V^0(\rho)\partial_\rho[b(\rho)\partial_\rho\psi_{V_n}]d\rho \, ,\nn\\
g_{A_n}&=&- \kappa\, \int_{{\rm D7 + \overline{D7}}}\psi_A^0(\rho)\partial_\rho[b(\rho)\partial_\rho\psi_{A_n}]d\rho \, .
\eea
Partially integrating the above, we have:
\be
g_n = - 2 \kappa \int_{\rho_0}^{\rho_{\Lambda}} \pd_{\rho} \left[ \psi^0 \,b(\rho) \,\pd_{\rho} \psi_{n} \right] + 2 \kappa \int_{\rho_0}^{\rho_{\Lambda}} (\pd_{\rho} \psi^0) \,b(\rho) \,(\pd_{\rho} \psi_{n}) \, .
\ee
Now, since $\pd_{\rho} \psi^0_V|_{q^2 = 0} = 0$, the vector decay constant $g_{V_n}$ is a total derivative. However, $\pd_{\rho} \psi^0_A|_{q^2 = 0} = \frac{1}{\rho_{\Lambda} - \rho_0} \neq 0$. Therefore, we find:
\be \label{gAn}
g_{A_n} = -2\kappa \,b(\rho) \,\partial_\rho\psi_{A_n}|_{\rho=\rho_{\Lambda}} + \frac{2\kappa}{\rho_{\Lambda} - \rho_0} \int_{\rho_0}^{\rho_{\Lambda}} b(\rho) \,\pd_{\rho} \psi_{A_n} \, ,
\ee
where we have used that $\psi^0_A|_{\rho = \rho_{\Lambda}} = 1$ and $\psi^0_A|_{\rho = \rho_0} = 0$. Now, partially integrating the last term in (\ref{gAn}) and using that $\psi_{A_n}|_{\rho = \rho_0, \,\rho_{\Lambda}} \!= 0$, we obtain:
\be
g_{A_n} = -2\kappa \,b(\rho) \,\partial_\rho\psi_{A_n}|_{\rho=\rho_{\Lambda}} - \frac{2\kappa}{\rho_{\Lambda} - \rho_0} \int_{\rho_0}^{\rho_{\Lambda}} (\pd_{\rho} b) \,\psi_{A_n} \, .
\ee
Since, to leading order in small $\beta$, the walking region has $b \approx const$, the second term vanishes; clearly, though, there will be subleading order corrections arising from that term. Thus, the exact expressions (\ref{coupling2}) are equivalent to 
\bea\label{coupling}
g_{V_n}&=&-2\kappa\, b(\rho)\partial_\rho\psi_{V_n}|_{\rho=\rho_{\Lambda}}\nn\\
g_{A_n}&=&-2\kappa\, b(\rho)\partial_\rho\psi_{A_n}|_{\rho=\rho_{\Lambda}} \, ,
\eea
that were used in \cite{HolTech}, {\it only in leading order} in small $\beta$.

To compute $g_n$ we need to know the properly normalized wave functions $\psi_n$, i.e. satisfying (\ref{norm}). To find them, we begin with the wavefunctions $\phi_n$ from Subsection \ref{MassSpec}, which should be normalized to $\delta_{n,m}$. Note that, in order to normalize $\phi_n$ to leading order in small $\beta$, we need only consider the region $z>z_{\bullet}$ since the contribution of the region $z<z_{\bullet}$ is suppressed by a power of $\beta$. Then we find from (\ref{phi2}):
\bea
\phi_{V_n}(z)&\simeq&\frac{1}{z_{\Lambda}J_1(r_n^J)}\sqrt{\,z}\,J_0(r_n^J\,z\,/\,z_{\Lambda}) \,,\nn\\
\phi_{A_n}(z)&\simeq&\frac{\sqrt{\,z}\,[J_0(m_n^A\,z)-R\, Y_0(m_n^A\,z)]}{z_{\Lambda}[J_1(\mu_n)-R\, Y_1(\mu_n)]\gamma_n}  \,,
\eea
where as before $R=J_0(\mu_n)\,/\,Y_0(\mu_n)$ and
\be\label{gam}
\gamma_n=\sqrt{\left|\left\{\frac{2}{\pi\,\mu_n[ J_1(\mu_n)-R\,Y_1(\mu_n)]}\right\}^2-1\right|} \,.
\ee
Now one can show that the wavefunctions, normalized as in (\ref{norm}), are given by:
\be\label{psin}
\psi_{V_n,A_n}(\rho)= 2^{3/2}\times 3^{1/8} c_0^{-3/8} \kappa^{-1/2} \beta^{1/16} e^{-\rho}\phi_{V_n,A_n}(z) \,.
\ee
Substituting (\ref{psin}) into (\ref{coupling}) we obtain:
\bea\label{coupling3}
g_{V_n}^2&=&\frac{4 \times3^{5/4}\kappa\,e^{-4\rho_\Lambda}(r_n^J)^2}{c_0^{3/4}\beta^{7/8}} \,,\nn\\
g_{A_n}^2&=&\frac{4\times3^{5/4}\beta^{1/8}\kappa\,e^{-4\rho_\Lambda}(\mu_n )^2}{c_0^{3/4}\beta^{7/8}}\gamma_n^{-2} \,.
\eea
Finally, from (\ref{sumrule}) we find for the $S$ parameter:
\be\label{S2}
S=\frac{\kappa\,c_0^{5/4}\pi\,\beta^{1/8}e^{4\rho_\Lambda}}{ 3^{3/4}}\sum_{n=1}^\infty\left[\frac{1}{(r_n^J)^2}-\gamma_n^{-2}\frac{1}{(\mu_n)^2}\right].
\ee
The sum in this expression can be evaluated numerically to any order $n$. 

\begin{figure}[t]
\begin{center}
\includegraphics[width=4in]{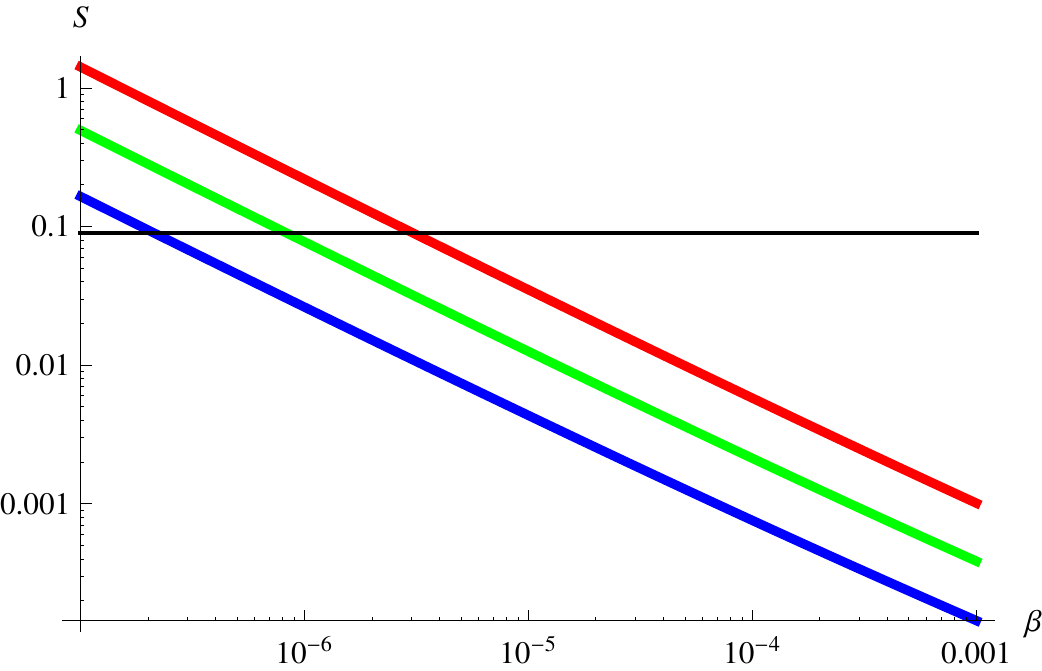}
\caption{The S-parameter as a function of $\beta$ at $e^{4(\rho_{\Lambda}-\rho_*)}=0.2$ (red), 0.065 (green), and 0.02 (blue) from (\ref{S1}) or (\ref{S2}). The black horizontal line represents the experimental upper limit.}
\label{spar}
\end{center}
\end{figure}

%\begin{figure}[htbp]
%\begin{center}
%\includegraphics[width=4in]{spar}
%\caption{The S-parameter as a function of $\beta$ at $e^{4(\rho_{\Lambda}-\rho_*)}=0.2$ (red), 0.065 (green), and 0.02 (blue) from (\ref{S1}) or (\ref{S2}). The black horizontal line represents the experimental upper limit.}
%\label{spar}
%\end{center}
%\end{figure}
Clearly, as in our earlier discussion regarding $\mu_n$, the numerical answer for $S$ also depends on the choice of cutoff. We found, however, that for every choice of $\rho_\Lambda$ the series in (\ref{S2}) converges rapidly. The $n=1$ term approximates the infinite sum within 20\%, while the sum of the first four terms, i.e. up to and including $n=4$, within 1\%. To illustrate the cutoff dependence, we plot in Fig.\ref{spar} \,the $S$ parameter as a function of $\beta$ (with $c_0=\sqrt{3}/16$ substituted) for three different values of $e^{4(\rho_{\Lambda}-\rho_*)}$ between $0.2$ and $0.02$. Note that there is a perfect agreement between the results in (\ref{S1}) and (\ref{S2}); the two curves coincide within 0.1\% for all $\beta$ and $\rho_{\Lambda}$, when we sum over the first 20 states.

\begin{figure}[t]
\begin{center}
\includegraphics[width=4in]{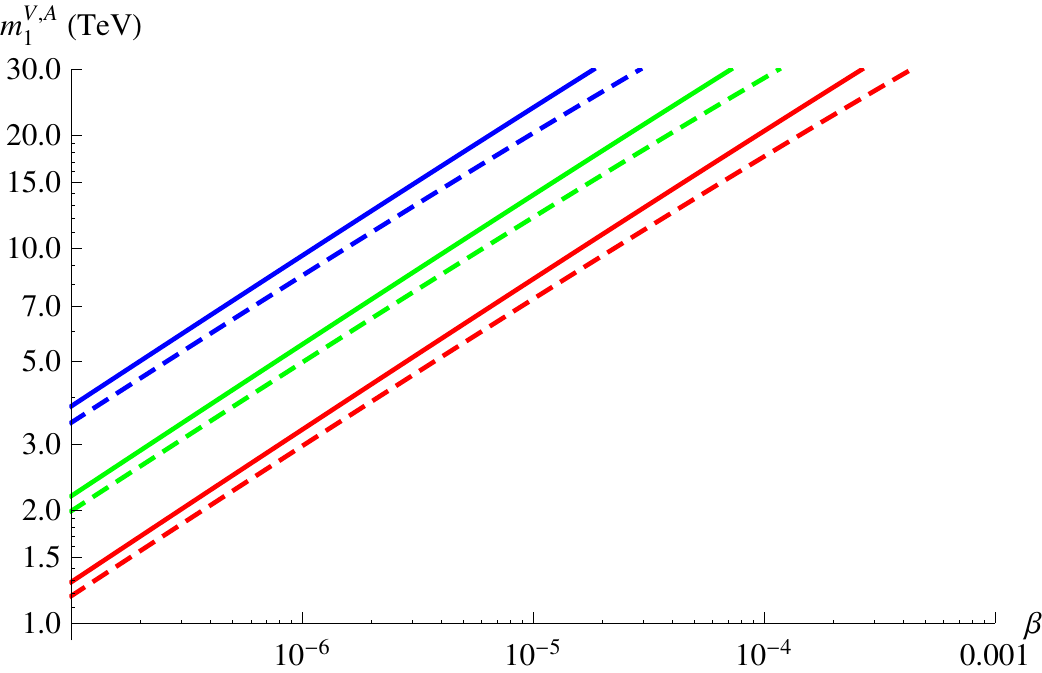}
\caption{The mass of the lightest vector (dashed lines) and axial vector (solid lines) bosons as a function of $\beta$ at $e^{4(\rho_\Lambda-\rho_*)}=0.2$ (red), $0.065$ (green) and $0.02$ (blue).}
\label{masses3}
\end{center}
\end{figure}

It is worth expressing the various physical quantities directly in terms of $F_{\pi}$.
From (\ref{fpi}), we can write:
\be\label{mkk}
M_{KK}=F_\pi \beta^{3/16}\sqrt{\frac{L}{\kappa}}\,(3\,c_0)^{-1/8} \, ,
\ee
which allows us to eliminate $M_{KK}$ from the expressions for the masses. Namely, substituting (\ref{mkk}) into (\ref{Vmass}), we obtain:
\be
m_n^V= e^{2(\rho_*-\rho_\Lambda)}\sqrt{\frac{6\,L}{\kappa}} \,\beta^{7/16}c_0^{-5/8}3^{-1/8} r_n^J F_\pi 
\ee
and similarly for $m_n^A$ with $r_n^J \rightarrow \mu_n$. Setting $F_\pi=246$ GeV, as appropriate for a technicolor model, we have plotted the dependence of the lightest vector and axial-vector masses on $\beta$ and $\rho_\Lambda-\rho_*$ in Fig.\ref{masses3}. Note that so far we have tacitly assumed that $N_{TF} = 1$, as in \cite{HolTech}. We should point out that increasing the number of flavors would decrease the masses of the mesons by a factor of $\sqrt{N_{TF}}$. 
%\begin{figure}[htbp]

Finally, let us also mention that one can relate directly the physical observables $S$, $F_{\pi}$ and $m_n^{V,A}$ to each other, without any parameters involved. Indeed, from (\ref{fpi}), (\ref{S1}), (\ref{Vmass}) and (\ref{Amass}), we obtain:
\bea\label{comb}
m_n^{V}&=&\sqrt{\frac{2\pi}{S}}F_\pi \,r_n^J \, ,\nn\\
m_n^{A}&=&\sqrt{\frac{2\pi}{S}}F_\pi \,\mu_n \, ,
\eea
where we have used the approximation $\sqrt{1-1/L}\to1$ with $L=\log(2 z_\Lambda^2\,/\,\lambda)\sim - \log(\beta)$. Note that the expression for $m_n^A$ is approximate since, due to (\ref{axialmass}), $\mu_n$ depends on $L$ as well. Setting $F_\pi=246$ GeV, we obtain for the mass of the lightest technirho boson:
\be\label{technirhomass}
m_1^V\simeq\frac{1}{\sqrt{S}}1.5 \,\mbox{ TeV}.
\ee
Hence the experimental upper bound for the $S$-parameter, i.e. $S\lesssim0.09$\,, implies the lower bound $m_1^V\gtrsim5$ TeV, for the case of a single flavor. If $N_{TF}>1$, then the lower bound on the mass of the lightest vector meson is smaller. We will investigate in more detail models with nonabelian flavor symmetry in a future publication.

\section{Discussion}

We studied a model of walking technicolor via its dual gravitational description. The latter was obtained by embedding probe D7-branes in a type IIB background. We were able to compute the electroweak $S$ parameter in two different ways. The first way is via an exact expression in terms of non-normalizable modes and the second one is via summing the contributions of the discrete vector and axial-vector states. The two answers are in perfect agreement, in contrast to previous holographic studies \cite{HolTech} of technicolor models. Hence, the issue that was raised there as an explanation for the discrepancy, namely the non-decoupling of KK modes, does not occur for us. %As a result, our effective description is more reliable. 

We found an unexpected dependence of the $S$ parameter on the length of the walking region. It would be very interesting to understand whether this is some kind of a characteristic prediction of stringy holographic (as opposed to the AdS/CFT-inspired phenomenological) walking models or a peculiarity of the particular model we studied here. To address this issue, one would have to try to enlarge the set of examples of gravitational duals of walking technicolor. One way to achieve that is to search for embeddings of the techniflavor probe branes, that are different compared to the one of \cite{LA}. Such embeddings would have to be more technically involved, but could potentially lead to interesting results. Another way to broaden the set of examples is to look for different walking backgrounds in string theory. At present, the only known ones are the background we considered here and a slight variant of it \cite{ModWalk} (see also \cite{EGNP}). They were found at the purely technical level, but at this point there is no good conceptual understanding of why they lead to walking. In order to obtain a wider class of solutions, it would be rather helpful to achieve a better understanding of those backgrounds. We hope to report on that in the future.

In this work, as well as in \cite{HolTech,LA}, the implicit assumption was that there is only one pair of probe techniflavor branes. If $N_{TF} > 1$, then our present results for $S$ and $F^2_{\pi}$ would be multiplied by a factor of $N_{TF}$. On the field theory side, this is just the usual statement that each technifermion makes an additive contribution to the spectral functions. On the gravity side, that we are studying here, this is due to the fact that for $N_{TF}$ probe D7-$\overline{{\rm D}7}$ pairs the total DBI action will be $N_{TF}$ times the action in (\ref{YM}). The presence of this factor on the right-hand side of (\ref{fpi}) would imply a decrease of the discrete spectrum masses by $\sqrt{N_{TF}}$. More importantly, in addition to the inclusion of an overall multiplier, the proper consideration of the $N_{TF} > 1$ case requires taking into account the nonabelian nature of the techniflavor field strength. This is crucial in order to be able to compute the $T$ and $U$ electroweak observables, which we will address in a future publication. 

The result we found for the S-parameter is positive-definite. However, although we can decrease it by varying the parameters of our model, we cannot reach $S=0$ within our approximations. Despite that, our result is below the lower bound that was conjectured in \cite{FS} based on the existence of an IR fixed point, the Banks-Zaks point \cite{BZ}. One should keep in mind, though, that our model is, quite likely, in a different universality class compared to the more familiar IR-fixed-point ones \cite{WalkTech}. A clear indication for this is provided by the fundamentally different role the number of techniflavours is playing in the two cases. Namely, in the IR fixed point considerations, it is crucial for the number of techniflavours to be comparable to the number of technicolors (i.e. $N_{TF} \sim N_{TC}$), in order to achieve a nearly conformal regime. In our considerations, on the other hand, one always has $N_{TF} <\!\!< N_{TC}$. Hence, our gravitational description is not a dual of a standard WTC model \cite{WalkTech}, but instead a dual of a novel kind of walking model. As a consequence, our positive-definite answer cannot be viewed as evidence against the possibility, considered in \cite{GT,AS}, of having models with negative $S$ parameter. In fact, it would be rather interesting to explore whether one could obtain $S < 0$ from a different kind of gravitational dual, probably with $N_{TF} \sim N_{TC}$. 

In order to reproduce on the gravity side the relation $N_{TF} \sim N_{TC}$, used in the IR fixed point models, one would have to go beyond the techniflavour probe approximation. The standard trick to construct a background with backreacted technfiflavour branes is to smear the latter along the transverse directions, in order to simplify the supergravity field equations. However, such a smearing is incompatible with the U-shaped embedding that realizes geometrically chiral symmetry breaking, which was of crucial importance for us. Hence, at present, finding a backreacted gravitational dual of walking technicolor, in which $N_{TF}$ is of the order of $N_{TC}$, is a technically-challenging open problem.

\section*{Acknowledgements}

We would like to thank C. Nunez for useful correspondence and P. Argyres, F. Sannino and R. Shrock  for valuable discussions. In addition, L.A. thanks the Simons Workshop in Mathematics and Physics, Stony Brook 2010, for hospitality during the initial stages of this work. Also, R.W. thanks the Aspen Center for Physics for hospitality, while working on this project. The research of L.A., P.S. and R.W. is supported by DOE grant FG02-84-ER40153.

\end{document}